\documentclass[twocolumn,english,prl]{revtex4-1}
\usepackage{ae,aecompl}
\usepackage{helvet}

\usepackage[T1]{fontenc}
\usepackage[latin9]{inputenc}
\usepackage{amsmath}
\usepackage{amssymb}
\usepackage{graphicx}

\makeatletter

\providecommand{\tabularnewline}{\\}

\usepackage{babel}
\usepackage{color}

\usepackage{babel}

\usepackage{babel}

\usepackage{babel}

\usepackage{babel}

\usepackage{babel}

\makeatother

\usepackage{babel}
\begin{document}

\title{Spectroscopic Evidence of the Aharonov-Casher effect in a Cooper
Pair Box}

\author{M.T. Bell$^{1,2}$, W. Zhang$^{1}$, L.B. Ioffe$^{1,3}$, and M.E.
Gershenson$^{1}$}

\address{$^{1}$Department of Physics and Astronomy, Rutgers University, 136
Frelinghuysen Rd., Piscataway, NJ 08854, USA}

\address{$^{2}$Department of Electrical Engineering, University of Massachusetts,
Boston, Massachusetts 02125}

\address{$^{3}$LPTHE, CNRS UMR 7589, 4 place Jussieu, 75252 Paris, France }
\begin{abstract}
We have observed the effect of the Aharonov-Casher (AC) interference
on the spectrum of a superconducting system containing a symmetric
Cooper pair box (CPB) and a large inductance. By varying the charge
$n_{g}$ induced on the CPB island, we observed oscillations of the
device spectrum with the period $\Delta n_{g}=2e$. These oscillations
are attributed to the charge-controlled AC interference between the
fluxon tunneling processes in the CPB Josephson junctions. The measured
phase and charge dependences of the frequencies of the $|0\rangle\rightarrow|1\rangle$
and $|0\rangle\rightarrow|2\rangle$ transitions are in good agreement
with our numerical simulations. Almost complete suppression of the
tunneling due to destructive interference has been observed for the
charge $n_{g}=e(2n+1)$. The CPB in this regime enables fluxon pairing,
which can be used for the development of parity-protected superconducting
qubits. 
\end{abstract}
\maketitle
The Aharonov-Casher (AC) effect is a non-local topological effect:
the wave function of a neutral particle with magnetic moment moving
in two dimensions around a charge acquires a phase shift proportional
to the charge \cite{PhysRevLett.53.319}. This effect has been observed
in experiments with neutrons, atoms, and solid-state semiconductor
systems (see, e.g., \cite{Cimmino1989,PhysRevLett.71.3641,PhysRevLett.96.076804}
and references therein). Similar effects have been predicted for superconducting
networks of nanoscale superconducting islands coupled by Josephson
junctions. For example, the wave function of the flux vortices (fluxons)
moving in such a network should acquire a phase that depends on the
charge on superconducting islands \cite{PhysRevD.40.4178}. Indeed,
oscillations of the network resistance in the flux-flow regime have
been observed as a function of the gate-induced island charge \cite{PhysRevLett.71.2311};
these oscillations have been attributed to the interference associated
with the AC phase. However, this attribution is not unambiguous, because
qualitatively similar phenomena can be produced by the Coulomb-blockade
effect due to the quantization of charge on the superconducting islands
\cite{PhysRevLett.88.050403}.

More recently, indirect evidence for the AC effect in superconducting
circuits has been obtained in the study of suppression of the macroscopic
phase coherence in one-dimensional (1D) chains of Josephson junctions
by quantum fluctuations \cite{PhysRevB.85.094503}. The quantum phase
slips (QPS) in the junctions can be viewed as the charge-sensitive
fluxon tunneling \cite{PhysRevLett.89.096802,Pop2010} provided the
conditions discussed below are satisfied. Microwave experiments \cite{PhysRevB.85.024521}
have demonstrated that dephasing of a fluxonium, a small Josephson
junction shunted by a 1D Josephson chain, can be due to the effect
of fluctuating charges on the QPS in the chain. Applications of the
AC effect in classical Josephson devices have been discussed in Refs.
\cite{PhysRevLett.88.050403,PhysRevLett.108.097001}.

In this Letter we describe microwave experiments which provide direct
evidence for the charge-dependent interference between the amplitudes
of fluxon tunneling. We have studied the microwave resonances of the
device consisting of two nominally identical Josephson junctions separated
by a nanoscale superconducting island (the so-called Cooper-pair box,
CPB) and a large inductance. A similar device with even greater kinetic
inductance provides a physical implementation of the fault tolerant
qubit (see below and Ref. \cite{Doucot2012}). The spectrum of the
device is determined by the QPS rate in the CPB junctions, which depends
on the charge of the superconducting island. The abrupt change of
the phase difference across each junction by $\sim2\pi$ (see below)
can be considered as adding/subtracting a single fluxon to the superconducting
loop formed by the CPB and the superinductor. We have observed almost
complete suppression of the fluxon tunneling due to the destructive
AC interference for the charge on the central CPB island $q=e(2n+1)$.
This complete suppression of fluxon tunneling provides an unequivocal
evidence for the Aharonov-Casher phase and clearly distinguishes this
effect from the Coulomb-blockade-related effects. Our results obtained
for this well-controlled system allow for direct quantitative comparison
with the theory.

\begin{figure}[bp]
\includegraphics[width=0.5\textwidth]{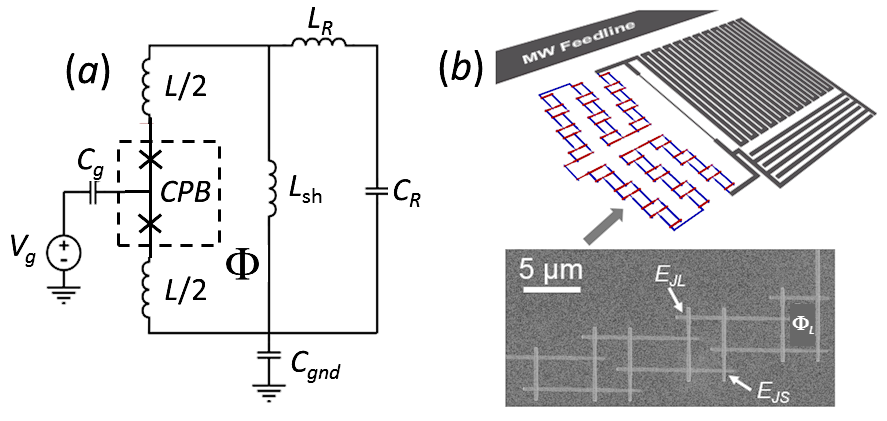} \protect\protect\protect\protect\protect\protect\protect\protect\caption{(color online) Panel (a): The schematics of the circuit containing
the device and the readout lumped-element resonator. The CPB Josephson
junctions are shown as crosses. Panel (b): The layout of the device,
the read-out resonator, and the MW transmission line. The superinductor
consists of 36 coupled cells, each cell represented a small superconducting
loop interrupted by three larger and one smaller Josephson junctions
\cite{PhysRevLett.109.137003}.}
\label{fig:fig1} 
\end{figure}

The studied device (Fig. \ref{fig:fig1}) consists of a superconducting
loop that includes a Cooper pair box and a superconducting inductor
with a large Josephson inductance $L$, the so-called superinductor
\cite{PhysRevB.85.024521}. Below we refer to this loop as the device
loop. The magnetic flux $\Phi$ in this loop controls the phase difference
across the superinductor. The design of our superinductor has been
described in Ref. \cite{PhysRevLett.109.137003}; the superinductor
used in this experiment consisted of 36 coupled cells, each cell represented
a small superconducting loop interrupted by three larger and one smaller
Josephson junctions (Fig. \ref{fig:fig1}b). The inductance $L$ reaches
its maximum when the unit cell is threaded by the magnetic flux $\Phi_{L}=\Phi_{0}/2$.
In this regime of full frustration, $L$ exceeds the Josephson inductance
of the CPB junctions by two orders of magnitude.

It is worth emphasizing that a large magnitude of $L$ and, thus,
a small value of the superinductor energy $E_{L}=(\frac{\Phi_{0}}{2\pi})^{2}\frac{1}{L}$,
is essential for the observation of the AC effect in our experiment.
Indeed, the classification of the device states by the discrete values
of the phase $\varphi=2\pi m$ and, thus, the notion of fluxons can
be justified if $E_{L}\ll E_{J}$ because only in this limit one can
ignore the phase drop across the CPB (for more details see Supplementary
Materials \cite{Materials}). In this respect, the studied device
resembles the fluxonium \cite{Manucharyan02102009}, in which a single
junction is shunted by a superinductor. Large inductance $L$ is an
important distinction of our device from the structure proposed in
Ref. \cite{PhysRevLett.88.050403} for the observation of suppression
of macroscopic quantum tunneling due to the AC effect. In the small-$L$
case considered in Ref. \cite{PhysRevLett.88.050403}, the phase weakly
fluctuates around the value $2\pi\frac{\Phi}{\Phi_{0}}$ and the phase
slips are completely suppressed (cf. Ref. \cite{Brink2002}). Note
that the condition $E_{L}\ll E_{J}$ was not satisfied in Ref. \cite{PhysRevB.85.094503},
so the data interpretation in terms of fluxon tunneling can be questioned.
Large $L$ values are also important for the spectroscopic measurements:
the superinductor reduces the device resonance frequency down to the
convenient-for-measurements 1-10 GHz range.

For the dispersive measurements of the device resonances, a narrow
portion of the device loop with the kinetic inductance $L_{sh}$ was
coupled to the read-out lumped-element resonator (for details of the
readout design, see \cite{PhysRevB.86.144512,PhysRevLett.112.167001}).
The global magnetic field, which determines the fluxes in both the
device loop, $\Phi$, and the unit cells of the superinductor, $\Phi_{L}$,
has been generated by a superconducting solenoid. The offset charge
on the CPB island was varied by the gate voltage $V_{g}$ applied
to the microstrip transmission line (Fig. \ref{fig:fig1}b).

\begin{table}[htp]
\begin{centering}
\protect\protect\protect\protect\protect\protect\protect\protect\caption{Parameters of Josephson junctions in the representative device. Parameters
of the CPB junctions correspond to the fitting parameters; parameters
of the superinductor junctions were estimated using the Ambegaokar-Baratoff
relationship and the resistance of the test junctions fabricated on
the same chip.}

\par\end{centering}

\begin{tabular}{|c|c|c|c|}
\hline 
Junctions  & In-plane areas, $\mu m^{2}$  & $E_{J},$GHz  & $E_{C},$GHz\tabularnewline
\hline 
CPB  & $0.11\times0.11$  & 6  & 6.4\tabularnewline
\hline 
Superinductor large  & $0.30\times0.30$  & 94  & 3.3\tabularnewline
\hline 
Superinductor small  & $0.16\times0.16$  & 25  & 11\tabularnewline
\hline 
\end{tabular}
\end{table}

\begin{figure}[tb]
\includegraphics[width=0.5\textwidth]{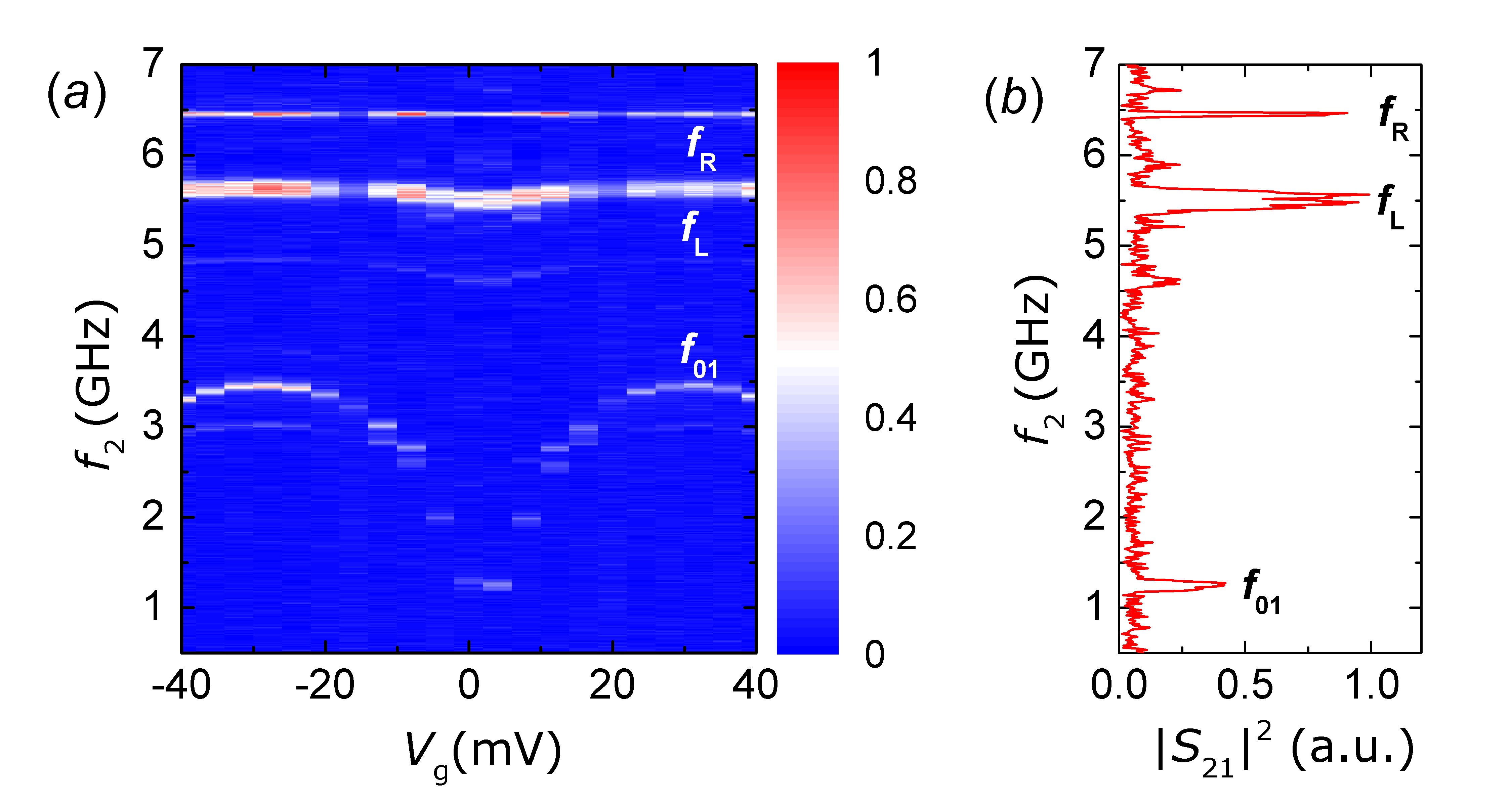} \protect\protect\protect\protect\protect\protect\protect\protect\protect\caption{(color online) Panel (a): The transmitted microwave power $|S_{21}|^{2}$
at the first-tone frequency $f_{1}$ as a function of the second-tone
frequency $f_{2}$ and the gate voltage $V_{g}$ measured at a fixed
value of $\Phi_{L}=0.5\Phi_{0}$. The power maxima correspond to the
resonance excitations of the device ($f_{2}=f_{01}$), the superinductor
($f_{L}$), and the read-out resonator ($f_{R}$). Note that the resonance
measurements could not be extended below $\sim1$ GHz because of a
high-pass filter in the second-tone feedline. Panel (b): The frequency
dependence of the transmitted microwave power measured at $V_{g}=0V$
and $\Phi_{SL}=0.5\Phi_{0}$. }
\label{fig:fig2-1} 
\end{figure}

The device, the readout circuits, and the microwave (MW) transmission
line (Fig. \ref{fig:fig1}b) were fabricated using multi-angle electron-beam
deposition of Aluminum through a lift-off mask (for fabrication details,
see Refs. \cite{PhysRevLett.112.167001,PhysRevB.86.144512}). Six
devices have been fabricated on the same chip; they were addressed
individually due to different resonance frequencies of the read-out
resonators. The parameters of the CPB junctions were nominally the
same for all six devices, whereas the maximum inductance of the superinductor
was systematically varied across six devices by changing the in-plane
dimensions of the small junctions in the superinductors \cite{PhysRevLett.109.137003}.
Below we discuss the data for one representative device; Table I summarizes
the parameters of junctions in the CPB junctions and superinductor
(throughout the Letter all energies are given in the frequency units,
1 K$\approx20.8$ GHz).

In the two-tone measurements, the microwaves at the second-tone frequency
$f_{2}$ excited the transitions between the $|0\rangle$ and $|1\rangle$
quantum states of the device, which resulted in a change of its impedance
\cite{Wallraff2004}. This change was registered as a shift of the
resonance of the readout resonator probed with microwaves at the frequency
$f_{1}$. The microwave set-up used for these measurements has been
described in Refs. \cite{PhysRevLett.109.137003,PhysRevB.86.144512,PhysRevLett.112.167001}.
The resonance frequency $f_{01}$ of the transition between the $|0\rangle$
and $|1\rangle$ states was measured as a function of the charge $n_{g}$
and the flux in the device loop. The $f_{01}$ measurements could
not be extended below $\sim1$ GHz because of a high-pass filter in
the second-tone feedline.

The results discussed below have been obtained in the magnetic fields
that correspond to $\Phi_{L}\thickapprox\Phi_{0}/2$ where $L$ reaches
its maximum \cite{PhysRevLett.109.137003}. Because the device loop
area ($\sim1,850\mu m^{2}$) was much greater than the superinductor
unit cell area ($15\mu m^{2}$), the phase across the chain could
be varied at an approximately constant value of $L$. All measurements
have been performed at $T=20$ mK.

\begin{figure}
\includegraphics[width=0.4\textwidth]{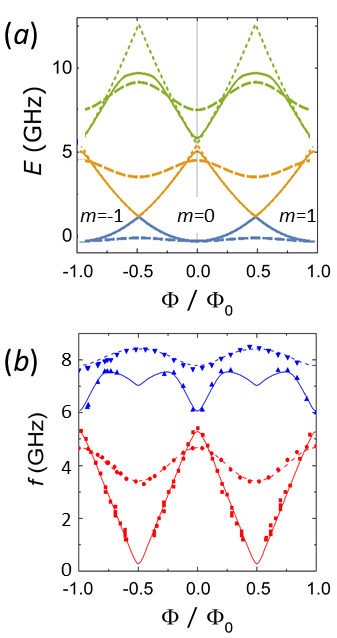}\protect\protect\protect\protect\protect\protect\protect\protect\caption{(color online) Panel (a): The flux dependence of the device energy
levels obtained by numerical diagonalization of the Hamiltonian (see
Supplementary Materials for details, the fitting parameters are listed
below). The solid curves correspond to $n_{g}=0.5$, the dashed curves
- to $n_{g}=0$ (the blue curves correspond to the ground state $|0\rangle$,
the yellow curves - to the state $|1\rangle$, and the green curves
- to the state $|2\rangle$). For comparison we also plotted the dotted
curves that correspond to the fully suppressed fluxon tunneling; in
this case there are no avoided crossings between the parabolas that
represent the superinductor energies $E_{L}(m,\Phi)=\frac{1}{2}E_{L}(m-\frac{\Phi}{\Phi_{0}})^{2}$
plotted for different $m$. Panel (b): The dependences of the resonance
frequencies $f_{01}$ (red dots - $n_{g}=0$, red squares - $n_{g}=0.5$)
and $f_{02}$ (blue down-triangles - $n_{g}=0$, blue up-triangles
- $n_{g}=0.5$ ) on the flux in the device loop. The theoretical fits
(solid curves - $n_{g}=0.5$, dashed curves - $n_{g}=0$) were calculated
with the following parameters: $E_{J}=6.25$ GHz, the asymmetry between
the CPB junctions $\triangle E_{J}=0.5$ GHz, $E_{C}=6.7$ GHz, $E_{L}=0.4$
GHz ($L=(\frac{\Phi_{0}}{2\pi})^{2}/E_{L}\eqsim0.4\mu H$), $E_{CL}=5$
GHz.}
\label{fig:fig3} 
\end{figure}

The resonances corresponding to the $|0\rangle\rightarrow|1\rangle$
transition are shown in Fig. \ref{fig:fig2-1}a as a function of the
gate voltage $V_{g}$ at a fixed value of the magnetic field that
is close to full frustration of the superinductor unit cells ($\Phi_{L}\backsimeq0.5\Phi_{0}$).
The dependence $f_{01}$ ($V_{g}$) is periodic in the charge on the
CPB island, $n_{g}$, with the period $\Delta n_{g}=1$ (here and
below the charge is measured in units $2e$ (mod $2e$)). The increase
of temperature above 0.3K resulted in reducing the period in half
due to the thermally generated quasiparticles population. Figure 2
also shows the resonance of the read-out resonator at $f_{R}=6.45$
GHz and the self-resonance of the superinductor $f_{L}\thickapprox5.5$
GHz. All three resonances are shown in Fig. 2b for $n_{g}\thickapprox0.47(V_{g}=0)$
and $\Phi_{L}\approx0.5\Phi_{0}$. Weaker resonances observed at $f_{2}\thickapprox3$
GHz and $4.8$ GHz at $V_{g}=-30mV$ correspond to the multi-photon
excitations of the higher modes of the superinductor.

Note that no disruption of periodicity neither by the quasiparticle
poisoning \cite{PhysRevLett.92.066802} nor by long-term shifts of
the offset charge was observed in the data in Fig. 2a that were measured
over 80 min. With respect to the quasiparticle poisoning, this suggests
that on average, the parity of quasiparticles on the CPB island remains
the same on this time scale. In the opposite case, the so-called ``eye\textquotedblright{}
patterns would be observed on the dependences of the resonance frequency
on the gate voltage \cite{Sun2012}. Significant suppression of quasiparticle
poisoning was achieved due to the gap engineering \cite{PhysRevLett.92.066802}
(the superconducting gap in the thin CPB island exceeded that of the
thicker leads by $\sim0.2$K), as well as shielding of the device
from infrared photons \cite{Barends2011}.

The expected flux dependence of the energy levels of the device is
shown in Fig. 3a. This flux dependence can be understood by noting
that in the absence of fluxon tunneling (the dotted curves in Fig.
3a corresponding to $n_{g}=0.5$ and identical CPB junctions) different
states are characterized by a different number $m$ of fluxons in
the device loop. At $E_{J}\gg E_{L}$ the energies of these states
are represented by crossing parabolas $E_{L}(m,\Phi)=\frac{1}{2}E_{L}(m-\frac{\Phi}{\Phi_{0}})^{2}$.
The phase slip processes mix the states with different numbers of
fluxons and lead to the level repulsion. The qualitative picture of
fluxon tunneling and AC interference is in good agreement with the
observed level structure shown in Fig. 3b.

Figure 3b shows the main result of this Letter: the dependences of
the resonance frequencies of the $|0\rangle\rightarrow|1\rangle$
and $|0\rangle\rightarrow|2\rangle$ transitions ($f_{01}$ and $f_{02}$,
respectively) on the flux in the device loop for the charges $n_{g}=0$
and $0.5$. In line with the level modeling, at $n_{g}=0$ the frequency
$f_{01}$ periodically varies as a function of phase, but never approaches
zero. On the other hand, when $n_{g}=0.5$, the amplitudes of fluxon
tunneling across the CPB junctions acquire the Aharonov-Casher phase
difference $\pi$. Provided that the CPB junctions are identical,
the destructive interference should completely suppress fluxon tunneling,
which results in vanishing coupling between the states $|m\rangle$
and $|m\pm1\rangle$ and disappearance of the avoided crossing. Since
the difference $E_{L}(m,\Phi)-E_{L}(m\pm1,\Phi)$ is linear in $\Phi$,
the spectrum at $n_{g}=0.5$ should acquire the sawtooth shape. This
is precisely what has been observed in our experiment. To better fit
the experimental data, we have assumed that the Josephson energies
are slightly different for the CPB junctions ($\triangle E_{J}<0.5$
GHz); for this reason, the minima of the theoretical sawtooth-shaped
dependence $f_{01}(\Phi)$ are slightly rounded. Fitting allowed us
to extract all relevant energies (see the caption to Fig. 3). The
amplitude of the single phase slips does not exceed 0.2 GHz, the amplitude
of the double phase slips is 0.4 GHz.

The studied device has the potential to become the building block
of the fault tolerant qubit. Namely, it can be used to implement a
protected qubit in which two logical states correspond to different
parities of fluxons in the device loop, the so-called ``flux-pairing''
qubit. Two conditions have to be satisfied for the realization of
protected states \cite{Doucot2012}. Firstly, the rate of cotunneling
of $pairs$ of fluxons should be significantly increased by reducing
the ratio $E_{J}/E_{C}$ for the CPB junctions. Note that at $n_{g}=0.5$,
the AC phase for cotunneling of fluxon pairs is $2\pi$ and the interference
is constructive. In this regime, the CPB represents a \textquotedbl{}$\cos(\phi/2)$\textquotedbl{}
Josephson element which energy is $4\pi$-periodic (see Supplementary
Materials \cite{Materials}). Secondly, for the proper operation of
the flux-pairing qubit, the inductance of the superinductor should
be further increased (approximately by an order of magnitude in comparison
with the device described above). To satisfy the latter challenging
requirement without reducing the superinductor resonance frequency,
the parasitic capacitance of the superinductor should be significantly
reduced. Such a qubit would not only be characterized by much improved
coherence, but, even more importantly, would enable certain fault-tolerant
gates \cite{Doucot2012}. The flux-pairing qubit is dual to a recently
realized charge-pairing qubit \cite{PhysRevLett.112.167001}.

To conclude, we have observed the effect of the Aharonov-Casher interference
on the spectrum of the Cooper pair box (CPB) shunted by a large inductance.
Large values of $L\;(E_{L}\ll E_{J}$) are essential for the observation
of the AC effect with the Cooper pair box; in this important respect
our devices differ from the earlier proposed structures \cite{PhysRevLett.88.050403}.
We have demonstrated that the amplitudes of the fluxon tunneling through
each of the CPB junctions acquire the relative phase that depends
on the CPB island charge $n_{g}$. In particular, the phase is equal
to $0\;(mod\;2\pi)$ at $n_{g}=2ne$ and $\pi\;(mod\;2\pi)$ at $n_{g}=e\left(2n+1\right)$.
The interference between these tunneling processes results in periodic
variations of the energy difference between the ground and first excited
states of the studied quantum circuit; the period of the oscillations
corresponds to $\Delta q=2e$. The phase slip approximation provides
quantitative description of the data and the observed interference
pattern evidences the quantum coherent dynamics of our large circuit.

We would like to thank B. Doucot for helpful discussions. The work
was supported in part by grants from the Templeton Foundation (40381),
the NSF (DMR-1006265), and ARO (W911NF-13-1-0431).

\section*{Supplementary Materials}

\section*{Energy spectra of the Cooper pair box shunted by a large inductance. }

The device studied in this paper consists of two very different elements:
a Cooper pair box (CPB) and a superinductor. The Cooper pair box is
described by the quantized value of the charge, or by a phase in the
interval $(0,2\pi)$. At significant CPB charging energy it is convenient
to use the former basis. In contrast, the superinductor is characterized
by continous conjugated variables, $\phi$(phase across) and $q$
(charge). The Cooper pair tunneling to the CPB island changes its
charge by $\pm1$; in the symmetric gauge the phase at the inductor
ends is $\pm\phi/2$, so the processes of tunneling from different
superinductor ends acquire phase factors $\exp(\pm\phi/2).$ Thus,
the full Hamiltonian describing the CPB coupled to the superinductor
is 
\begin{align}
H= & -E_{J}\left(a^{+}+a^{-}\right)\cos(\phi/2)+4E_{C}(n-n_{g})^{2}+4E_{CL}q^{2}\nonumber \\
 & +\frac{1}{2}E_{L}(\phi-2\pi\Phi/\Phi_{0})^{2}\label{eq:H}
\end{align}
where $a^{\pm}$are operators that raise (lower) the charge of the
CPB island, $n_{g}$ is the charge induced by the gate, and $E_{CL}$
is the effective charging energy of the superinductor that is due
to the capacitors of its junctions and ground capacitance of the whole
structure.

The analysis is further simplified for large charging energies $E_{C}\gg E_{J}$
which is marginally satisfied in the studied device ($E_{J}=0.29\,\mbox{K}$,
$E_{C}=0.31\,\mbox{K}$). In this case the Cooper pair tunneling is
significant only in the vicinity of the full charge frustration, $n_{g}=N+0.5$
where the only relevant charging states are $n=N,\,N+1$. In the reduced
space Hamiltonian becomes 
\begin{align}
H_{R}= & -E_{J}\cos(\phi/2)\sigma^{x}+4E_{C}(n_{g}-0.5)\sigma^{z}+4E_{CL}q^{2}\nonumber \\
 & +\frac{1}{2}E_{L}(\phi-2\pi\Phi/\Phi_{0})^{2}\label{eq:H_R}
\end{align}
Away from the charge frustration ($n_{g}=0.5$) the charge fluctuations
are small, so that $\sigma^{z}\approx-1$ in all low energy states.
Duality between phase and charge implies that in this situation the
phase fluctuations are large. In this approximation the low energy
states coincide with those of the oscillator with $\omega_{0}=\sqrt{8E_{CL}E_{L}}.$
Charge fluctuations lead to a weak flux dependence of the energies
of these states. In the leadling order to the perturbation theory
the oscillator potential becomes 
\[
V=\frac{-E_{J}^{2}}{4E_{C}}\cos\phi+\frac{1}{2}E_{L}(\phi-2\pi\Phi/\Phi_{0})^{2}
\]
that leads to the weak dependence of the oscillator level on the flux
$E_{01}=\omega_{0}+\delta E\cos2\pi\Phi/\Phi_{0}.$ This dependence
is exactly the one observed experimentally (red data points in Fig.
3b).

In the opposite limit, close to the full frustration, at $\left|n_{g}-0.5\right|\ll E_{J}/E_{C}$,
the phase slip amplitude vanishes due to the distructive Aharonov-Casher
interference. Formally in this limit one can treat the second term
in (\ref{eq:H_R}) as perturbation. In the zeroth order of the perturbation
theory one obtains two independent sectors characterized by $\sigma^{x}=\pm1$
that we shall refer to as even/odd sectors below. Exactly at $n_{g}=0.5$
these sectors are completely independent. At $E_{J}\gtrsim E_{CL}\gg E_{L}$
in each sector the low energy wavefunctions are localized in the vicinity
of points $\phi=2\pi m$, $m$ being the number of fluxons in the
phase loop. The energy spectrum in this case is given by the set of
parabolas shown in Fig. 3a. In this approximation the energy levels
corresponding to different parabolas cross. The level repulsion between
the levels represented by adjacent parabolas is due to the phase slips
that vanish at $n_{g}=0.5$.

The level repulsion between the levels represented by next nearest
parabolas is due to the double phase slips. Formally it is decribed
by the quantum tunneling in the effective potential

\begin{equation}
V(\phi)=\pm E_{J}\cos(\phi/2)+\frac{1}{2}E_{L}(\phi-2\pi\Phi/\Phi_{0})^{2}\label{eq:pot}
\end{equation}

It occurs with amplitude 
\begin{equation}
t=A(g)g^{1/2}\exp(-g)\omega_{p}\label{eq:t}
\end{equation}
where $g=4\sqrt{2E_{J}/E_{CL}}$, $\omega_{p}=\sqrt{2E_{J}E_{CL}}$
and $A(g)\approx0.8$. This tunneling process changes $m\rightarrow m\pm2$,
but does not mix even and odd sectors. In the limit of large $E_{J}\gg E_{CL}$
the amplitude becomes exponentially small which was the case for the
studied device ($g\approx6$). In this case the transitions due to
double phase slips are almost completely suppressed. The energies
of the states with different $m$ are quadratic as a function of the
flux, exactly at half flux quantum these energies cross. So the energy
difference between the ground and first excited states is linear in
$\Phi$ with zero intercept. This is exactly the behavior observed
in the data (Fig. 3b).

Note that the classification of states by the discrete values of the
phase is only possible if $E_{L}\ll E_{J}.$ In the opposite limit
$E_{L}\geq E_{J}$ the phase experiences small oscillations around
$2\pi\Phi/\Phi_{0}$ and the phase slips are completely suppressed.
In the intermediate regime the phase is localized around the minima
of the potential energy $V(\phi)$ that differ from each other by
a non-integer fraction of $2\pi$.

The linearity of $E_{01}(\Phi)$ at $n_{G}=0.5$ is disturbed by two
factors. The first one is the difference in $E_{J}$ of two junctions
comprising the CPB. Due to this difference the flux tunneling is not
completely suppressed even at $n_{g}=0.5$. This leads to the level
repulsion at half integer flux at which neighboring parabolas intersect.
This would lead to some rounding of $E_{01}(\Phi)$ around half integer
flux. Within the experimental accuracy, the data do not show such
effect which implies that the difference between two $E_{J}$ is small.
The second factor is more interesting, it is due to a significant
tunneling rate of two fluxons through the CPB. This leads to the level
repulsion between the levels corresponding to next neighboring parabolas
in Fig. 3a and rounding of the maxima of the spectra data at $n_{g}=0.5$
(blue points in Fig. 3b). Some hint of this behavior can be seen in
the data, we estimate that $t\approx0.2\,\mbox{GHz}$.

The analytical results obtained above become quantitatively correct
in the regime $E_{C}\gg E_{J}$ but they remain qualitatively correct
even for $E_{C}\gtrsim E_{J}.$ For more precise quantitative description
of the experiment in this regime we used numerical diagonalization
of the full Hamiltonian (\ref{eq:H}). This allowed us to obtain the
spectra for all values of the induced charges, $n_{g},$ and unambigous
data fit. We found that for the experimental device parameters it
is sufficient to restrict oneself to the three lowest energy charging
states and range of $\phi=(-12\pi,12\pi)$. In this approximation
the Hamiltonian becomes $3M\times3M$ matrix where $M$ is the number
of discrete values that were used to approximate the continuos variable
$\phi.$ Very accurate results can be achived by using $0.2\pi$ increments
(i.e. 20 steps for $4\pi$ period).

\section*{Quantum state protection expected for large inductance. }

We now briefly discuss the behavior expected for significant double
flux tunneling and very large inductance, at which, as we now show,
one expects protection against both the charge and flux noise. Generally,
tunneling between the states with different $m$ implies that the
full wave function is the superposition of the states with different
$m$ that can be found from the diagonalization of the discrete oscillator
Hamiltonian 
\begin{align}
H_{o}= & -t\left(\left|m\right\rangle \left\langle m+2\right|+\left|m+2\right\rangle \left\langle m\right|\right)+\nonumber \\
 & 2\pi^{2}E_{L}(m-\frac{\Phi}{\Phi_{0}})^{2}\label{eq:H_o}
\end{align}
where $m$ corresponds to either even or odd numbers. At large $t\gg E_{L}$
the low energy states in each sector are spead over many different
$m$ and are almost intistinguishable. This implies the protection
against the external noises. More quantitatively, the dependence on
$m_{0}$ is given by 
\begin{eqnarray}
E(m_{0}) & = & 2A(G)G^{1/2}\exp(-G)\Omega\cos(2\pi m_{0})\label{eq:E(m_0)}\\
G & =\frac{2}{\pi} & \sqrt{\frac{2t}{E_{L}}}\label{eq:G}\\
\Omega & = & 4\pi\sqrt{2tE_{L}}\label{eq:Omega}
\end{eqnarray}
This dependence becomes exponentially small at $G\gg1$ that indicates
weak sensitivity to external flux.

The charge noise can be described as $n_{g}(t)$ variations. Non-zero
value of $n_{g}-0.5$ results in a small amplitude that mixes even
and odd sectors. This amplitude is given by 
\begin{equation}
t_{eo}\approx\pi^{3/4}g^{1/4}E_{C}(n_{g}-0.5)\sqrt{\frac{t}{\omega_{p}}}\label{eq:t_m}
\end{equation}
Exactly at $\Phi=\Phi_{0}/2$ the energies of the odd and even sectors
become equal in the absence of $t_{eo}.$ Non-zero $t_{eo}$ leads
to level splitting but the effect is small in $n_{g}-0.5$ and $(t/\omega_{p})$.

The equations (\ref{eq:E(m_0)}-\ref{eq:t_m}) allow one to derive
the conditions for optimal charge and flux protection. Generally,
the coupling to the charge noise is largest at $\Phi=\Phi_{0}/2$
because $E_{01}=2t_{m}\sim(n_{g}-0.5)$ but this coupling becomes
small at $t\ll\omega_{p}.$ The flux noise affects the energy levels
via $m_{0}(\Phi)$ dependence. This effect becomes small at $t\gg E_{L}$.
Thus, the optimal protection against the noises is achieved for $\omega_{p}\gg t\gg E_{L}.$
Notice that while the first inequality is easy to achieve, the second
requires a large superinductance. For instance in this experiment
$t/E_{L}\approx0.05-0.5$. For significant protection against flux
noise $t/E_{L}$ should be greater than 10, while for the good charge
noise protection one needs $\omega_{p}/t\gtrsim10^{2}$ that results
in the condition $\omega_{p}/E_{L}\gtrsim10^{3}.$

\expandafter\ifx\csname natexlab\endcsname\relax\global\long\def\natexlab#1{#1}
 \fi \expandafter\ifx\csname bibnamefont\endcsname\relax \global\long\def\bibnamefont#1{#1}
 \fi \expandafter\ifx\csname bibfnamefont\endcsname\relax \global\long\def\bibfnamefont#1{#1}
 \fi \expandafter\ifx\csname citenamefont\endcsname\relax \global\long\def\citenamefont#1{#1}
 \fi \expandafter\ifx\csname url\endcsname\relax \global\long\def\url#1{\texttt{#1}}
 \fi \expandafter\ifx\csname urlprefix\endcsname\relax\global\long\def\urlprefix{URL }
 \fi \providecommand{\bibinfo}[2]{#2}

\end{document}